\documentclass[a4paper,11pt]{article}
\pdfoutput=1 

\usepackage{jheppub} 

\usepackage[T1]{fontenc} 
\usepackage{multirow}%

\title{\boldmath Search for sub-millicharged particles at J-PARC}

\author[a]{Jeong Hwa Kim}
\author[a]{In Sung Hwang}
\author[a,1]{Jae Hyeok Yoo, \note{Corresponding author.}} 

\affiliation[a]{Korea University, \\145 Anam-ro, Seongbuk-gu, Seoul, 02841, Korea}

\emailAdd{jailbraker@korea.ac.kr}
\emailAdd{his5624@korea.ac.kr}
\emailAdd{jaehyeokyoo@korea.ac.kr}

\abstract{
  We studied the feasibility of an experiment searching for sub-millicharged particles ($\chi$s) using 30 GeV proton fixed-target collisions at J-PARC. The detector is composed of two layers of stacked scintillator bars and PMTs and is proposed to be installed 280 m from the target. The main background is a random coincidence between two layers due to dark counts in PMTs, which can be reduced to a negligible level using the timing of the proton beam. With $N_\textrm{POT}=10^{22}$ which corresponds to running the experiment for three years, the experiment provides sensitivity to $\chi$s with the charge down to $5\times10^{-5}$ in $m_\chi<0.2$ $\textrm{GeV}/\textrm{c}^2$ and $8\times10^{-4}$ in $m_\chi<1.6$ $\textrm{GeV}/\textrm{c}^2$. This is the regime largely uncovered by the previous experiments. We also explored a few detector designs to achieve an optimal sensitivity to $\chi$s. The photoelectron yield is the main driver, but the sensitivity does not have a strong dependence on the detector configuration in the sub-millicharge regime.
}

\begin{document} 
\maketitle
\flushbottom

\section{Introduction}

Electric charge quantization is a long-standing question in particle physics. While Grand Unified Theories (GUTs) have typically been thought to preclude the possibility for particles that do not have integer multiple electron charge (millicharged particles hereafter), well-motivated dark-sector models~\cite{ArkaniHamed:2008qn,Pospelov:2008jd} have been proposed to predict the existence of millicharged particles while preserving the possibility for unification. Such models can contain a rich internal structure, providing candidate particles for dark matter. Recent results from the EDGES experiment~\cite{EDGES1} suggest that the observed 21-cm absorption profile can be explained if a fraction of dark matter is composed of millicharged particles~\cite{EDGES2}. 

One well-motivated mechanism that leads to millicharged particles is to introduce a new $U(1)$ in the dark sector with a massless dark-photon and a massive dark-fermion ($\chi$)~\cite{Holdom:1985ag,Izaguirre:2015eya}. In this scenario, the dark-photon and the photon in the Standard Model kinematically mix and the charge of $\chi$ is determined by the size of the mixing. Therefore, depending on the strength of mixing, $\chi$ can have an electric charge that is not integer multiple. Hereafter, $\chi$ is used to denote millicharged particles.

A number of experiments have searched for millicharged particles, including in an electron fixed-target experiment~\cite{MilliQ}, proton-proton colliders~\cite{Chatrchyan_2013,Chatrchyan_2013_2,PhysRevD.102.032002}, proton fixed-target experiment~\cite{bebc} and neutrino experiments~\cite{Davidson:1991si,PhysRevLett.124.131801}. A comprehensive review is in Reference~\cite{Emken_2019}. In the parameter space of the charge ($Q$) and mass ($m_\chi$), the region of $m_\chi>0.1$ $\textrm{GeV}/\textrm{c}^2$ and $Q<10^{-2}e$ is largely unexplored.

Proton fixed-target experiments provide a solid testing ground for $\chi$s. The particle flux is much larger than the collider experiments and they can reach a higher energy regime than electron fixed-target experiments. The sensitivity of such experiments to $\chi$s can reach beyond $Q\sim 10^{-3}e$ for a wide mass range from a few $\textrm{MeV}/\textrm{c}^2$ to a few $\textrm{GeV}/\textrm{c}^2$. This letter proposes a new experiment, SUBMET (SUB-Millicharge ExperimenT), which utilizes the $30$ GeV proton beam at Japan Proton Accelerator Research Complex (J-PARC) to search for $\chi$s in this unexplored region.

\section{Production of millicharged particles at J-PARC}

At proton fixed-target collisions at J-PARC, $\chi$s with charge $Q$ can be produced from the decay of $\pi^0, \eta$ and $J/\psi$ neutral mesons. The $\Upsilon$ production is not relevant because the center-of-mass energy is $7.5$ GeV for the collisions between the $30$ GeV proton beam and the fixed target. The lighter mesons ($\mathfrak{m} = \pi^0,\eta$) decay through photons ($\pi^0, \eta \to \gamma\chi\bar{\chi}$), while the $J/\psi$ decays to a pair of $\chi$s directly ($J/\psi \to \chi\bar{\chi}$). In both cases, $m_\chi$ up to $m_\mathfrak{m}/2$ is kinematically allowed. The number of produced $\chi$s ($N_\chi$) can be calculated by the equation in~\cite{fermini},
\begin{eqnarray}
N_\chi \propto c_\mathfrak{m} \epsilon^2 N_\textrm{POT} \times f\left(\frac{m^2_\chi}{m^2_\mathfrak{m}}\right)
\end{eqnarray}
where $c_\mathfrak{m}$ is the number of mesons produced per proton-on-target (POT), $N_\textrm{POT}$ is the total number of POT, $\epsilon = Q/e$, and $f$ is a phase space related integral. The $c_\mathfrak{m}$ of each meson is extracted using \texttt{PYTHIA8}~\cite{Sjostrand:2014zea} and the estimated values are $c_{\pi^0}=1.9$, $c_{\eta}=0.21$, and $c_{J/\psi}=5\times10^{-9}$. Assuming $N_\textrm{POT}=10^{22}$ that corresponds to running the experiment for $3$ years~\cite{Oyama:2020kev}, the expected number of $\chi$s that reach the detector is in the order of $10^{16}$ at $\epsilon=1$ and $10^{9}$ at $\epsilon=10^{-4}$.

\section{Experimental site and detector concept}
\label{sec:site_detector}

\begin{figure}[h]
\begin{center}
\includegraphics[width=0.99\linewidth]{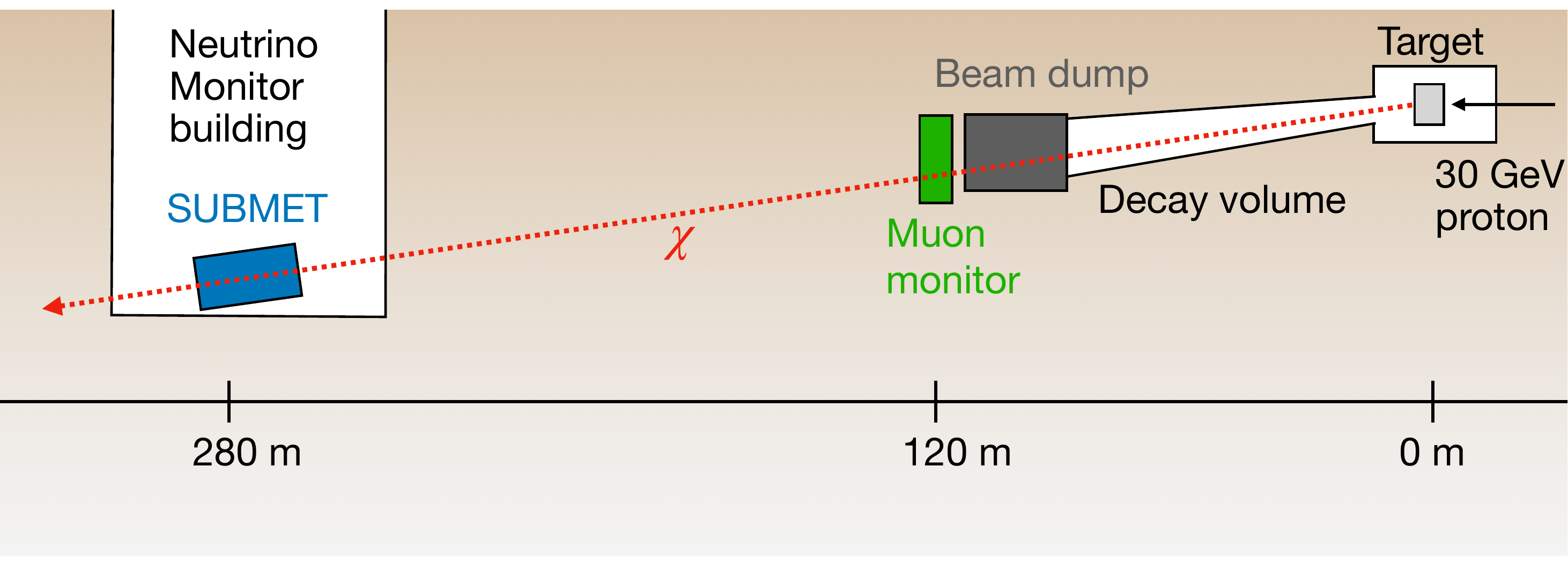}
\end{center}
\caption{Illustration of the experimental site. $\chi$s are produced near the target and reach SUBMET after penetrating the beam dump, the muon monitor and the sand. The detector is located $280$ m from the target and approximately $30$ m underground.}
\label{fig:site}
\end{figure}

In J-PARC a $30$ GeV proton beam is incident on a graphite target to produce hadrons that subsequently decay to a pair of muon and muon neutrino in the decay volume. The remaining hadrons are then dumped in the beam dump facility. Since they are Minimum Ionizing Particles, muons can penetrate the beam dump and be identified by the muon monitor located behind the beam dump facility. The on-axis near detector, Interactive Neutrino GRID (INGRID)~\cite{t2k}, is inside the Neutrino Monitor (NM) building located $280$ m from the target. The space between the muon monitor and INGRID is filled with sand. The experimental site is illustrated in Figure~\ref{fig:site}. The proton beam has a repetition rate of 1.16 s and each spill contains 8 bunches which are separated by $600$ ns~\cite{Friend_2017}. The beam timing is available at the site and this allows for substantial suppression of backgrounds at the level of $O(10^{-6})$. 

If $\chi$s are produced, they penetrate the space between the target and the detector without a significant energy loss because of their feeble interaction with matter. Therefore, they can be detected at the NM building if a detector sensitive to identifying such particles is installed. The area behind the V-INGRID on B2 ($\sim30$ m underground) is unoccupied and can be a potential detector site. The distance from the axis of the neutrino beam is $\sim 5$ m. 

\begin{figure}[h]
\begin{center}
\includegraphics[width=0.8\linewidth]{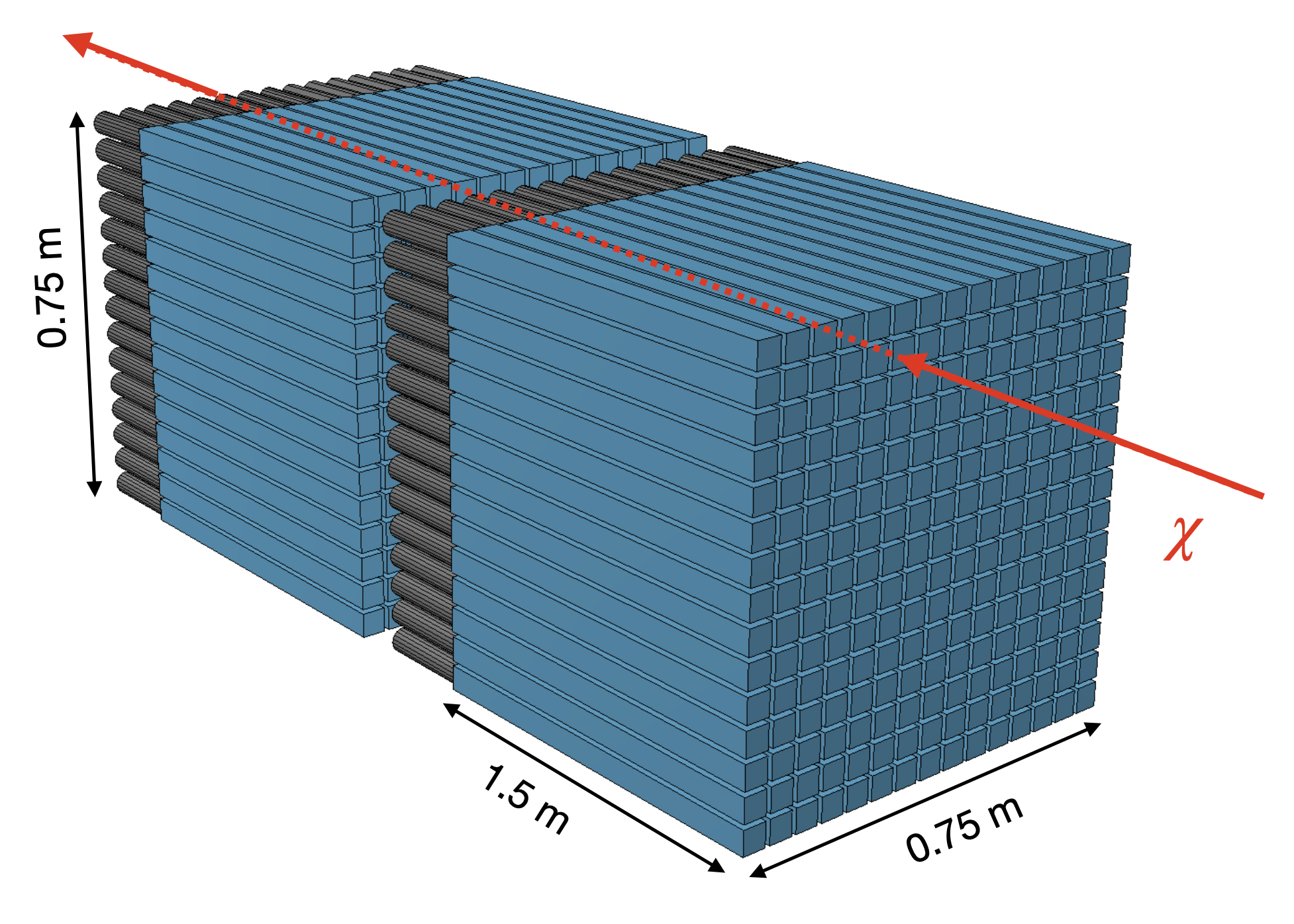}
\end{center}
\caption{Demonstration of the SUBMET detector. There are two layers of stacked scintillator bars (blue). At one end of each bar, a PMT (black) is attached. $\chi$s penetrate both layers in a narrow time window.}
\label{fig:detector}
\end{figure}

The detector concept proposed for this experiment is based on a similar proposal made in~\cite{Ball:2016zrp}, sharing the idea to use a segmented detector with large scintillator bars. To be sensitive to charges below $10^{-3}e$, a thick sensitive volume is needed. It is advantageous to segment the large volume because it helps reducing backgrounds due to dark currents and shower particles from cosmogenic muons to a negligible level. It also allows for utilizing the directionality of the incident $\chi$s to further suppress non-pointing particles. The detector, as shown in Figure~\ref{fig:detector}, is composed of 2 layers of stacked $5\times5\times150$ $\textrm{cm}^3$ BC-408 plastic scintillator bars~\cite{bc408}. They are aligned such that the produced $\chi$s pass through both layers in a narrow time window. In each layer there are $15 \times 15$ scintillator bars, so the area of the detector face is about $0.5 \textrm{ m}^2$. A prototype of a detector with a similar design has been installed at the LHC, and shown robustness and sensitivity to $\chi$s~\cite{PhysRevD.102.032002}.

At the end of each scintillator bar, a photodetector is attached to convert the photons to an electronic signal. Photomultipliers (PMTs) are suitable for this experiment because of their large area coverage, low cost, and low dark current. The total volume of the detector is approximately $0.75\times0.75\times3.5$ $\textrm{m}^3$ including the PMTs.

The signal acceptance rate, the fraction of $\chi$s that go into the detector area of $0.5\textrm{ m}^2$ at $280$ m from the target, is calculated as a function of distance from the beam axis to the detector. It is in the order of $O(10^{-4})$ and does not depend on the position strongly, up to a few meters from the axis, since the detector is located far from the target. At 5 m the rate is only $\sim 10$\% lower than the on-axis region. This provides some flexibility in selecting the location of the detector. The effect of energy loss and multiple Coulomb scattering in the sand is estimated to be negligible for the charge range of interest, particularly below $10^{-3}e$, so they have a small impact on the sensitivity of the experiment.

\section{Background sources}

$\chi$s that reach the detector will go through both layers within a $\sim10$ ns time window producing a coincidence signal. In this section, the background sources that can mimic this coincidence signal are discussed. They can be divided into three categories; random coincidence, beam-induced, and cosmic-induced backgrounds. 

In PMTs, spurious current pulses can be produced by thermal electrons liberated from the photocathode. Therefore, a random coincidence of such pulses in different layers can be identified as a millicharge signal. The typical size of the pulses is very small and this makes random coincidence the major background source in $Q<10^{-3}e$ regime. The rate of random coincidence can be large depending on the rate of the spurious pulses (dark count rate, DCR) even if the time window for the coincidence signal is $10$ ns. The random coincidence rate is $n N^n \tau^{n-1}$ where $n$ is the number of layers, $N$ is the DCR, and $\tau$ is the coincidence time window. Using a typical PMT DCR of $500$ Hz at room temperature, $n=2$, and $\tau=10$ ns, the random coincidence rate of two bars is $0.15$ per year. There can be $15\times15=225$ such coincidence signals, so the total coincidence rate is $\sim 35$ per year. The liberation of electrons is a thermal activity, which can be reduced by cooling the cathodes. With $N=100$ Hz, the random coincidence background is reduced to $1.5$ events per year. 

Muons are produced from the pion decays in the decay volume together with neutrinos. The density of quartz, which typically takes up the largest fraction of sand, is $2.65$ $\textrm{g}/\textrm{cm}^3$ and $dE/dx = 1.699 \textrm{ MeV}\textrm{cm}^2/\textrm{g}$~\cite{Groom:2001kq}, so the energy loss of a MIP in $>100$ m of sand is much larger than $30$ GeV. Therefore, such beam-induced muons can't reach the detector. Although the muons from the pion decays can't reach the detector, neutrinos can and may interact with the scintillator material to produce small signals. The number of neutrino interaction events in INGRID is $\sim 1.5 \times 10^8$ for $N_\textrm{POT}=10^{22}$~\cite{Abe:2011xv}. Since a large fraction of INGRID material is iron, the rate of neutrino interaction in INGRID can be used as an upper bound for SUBMET. One layer of SUBMET is approximately $30$ times smaller, so the rate is $\sim 5 \times 10^6$ for $N_\textrm{POT}=10^{22}$ in one layer of SUBMET. Requiring coincidence in two layers, the expected number of this background becomes negligible. The interaction of the neutrinos and the material of the wall of the NM building in front of the detector can produce muons that go through the detector. These muons can be identified and rejected by installing scintillator plates between the wall and the detector or by using the very large scintillation yield of a muon that can be separated from the millicharge signal.

Cosmic muons that penetrate the cavern or the materials above the detector can produce a shower of particles that is large enough to hit both layers simultaneously. In such events, the hits in multiple layers can be within the coincidence time window and will look like a signal event. The particles in the shower generate more photons than $\chi$s, so the signals from cosmic muon showers can be rejected by vetoing large pulses. As done to tag the muons produced in the wall of the NM building in front of the detector, scintillator plates can be installed covering the whole detector to tag any ordinary-charged particles or photons incident from top and sides of the detector. These auxiliary components were proven to be effective in rejecting events with such particles~\cite{PhysRevD.102.032002}. In addition, the cosmic shower penetrates the detector sideways, leaving hits in multiple bars in the same layer, while $\chi$s will cause a smaller number of hits. A cosmic shower and signals from radioactive decays overlapping with dark current can be another source of background. Since the rate of this background depends on the environment strongly, a precise measurement can be performed \textit{in situ} only.

To estimate the sensitivity of the experiment, we assume that the total background ($N_\textrm{bkg}$) over three years of running is $5$ events.

\section{Sensitivity}

The probability of detecting a $\chi$ in an $n$-layer detector is given by Poisson distribution $P = (1-e^{-N_\textrm{PE}})^n$ where $N_\textrm{PE}$ is the number of photoelectrons. $N_\textrm{PE}$ is proportional to the quantum efficiency (QE) of PMT, $\epsilon^2$, and the number of photons that reach the end of the scintillator ($N_\gamma$). The $\epsilon^2$ term comes from the fact that the energy loss of a charged particle in matter is proportional to $Q^2$. In order to calculate $N_\gamma$ a \texttt{GEANT4}~\cite{Agostinelli:2002hh} simulation is performed. Using a $5\times5\times150$ $\textrm{ cm}^3$ BC-408 scintillator with a surface reflectivity of 98\%,  $N_\gamma$ is $8.3 \times 10^5$. Taking QE into account, $N_\textrm{PE}$ is $2.5\times 10^5 \epsilon^2$. Once we have $N_\textrm{PE}$ and $P$, The total number of signal events measured by the detector can be calculated as $s = N_\textrm{PE} P$.

\begin{figure}[h]
\begin{center}
\includegraphics[width=0.99\linewidth]{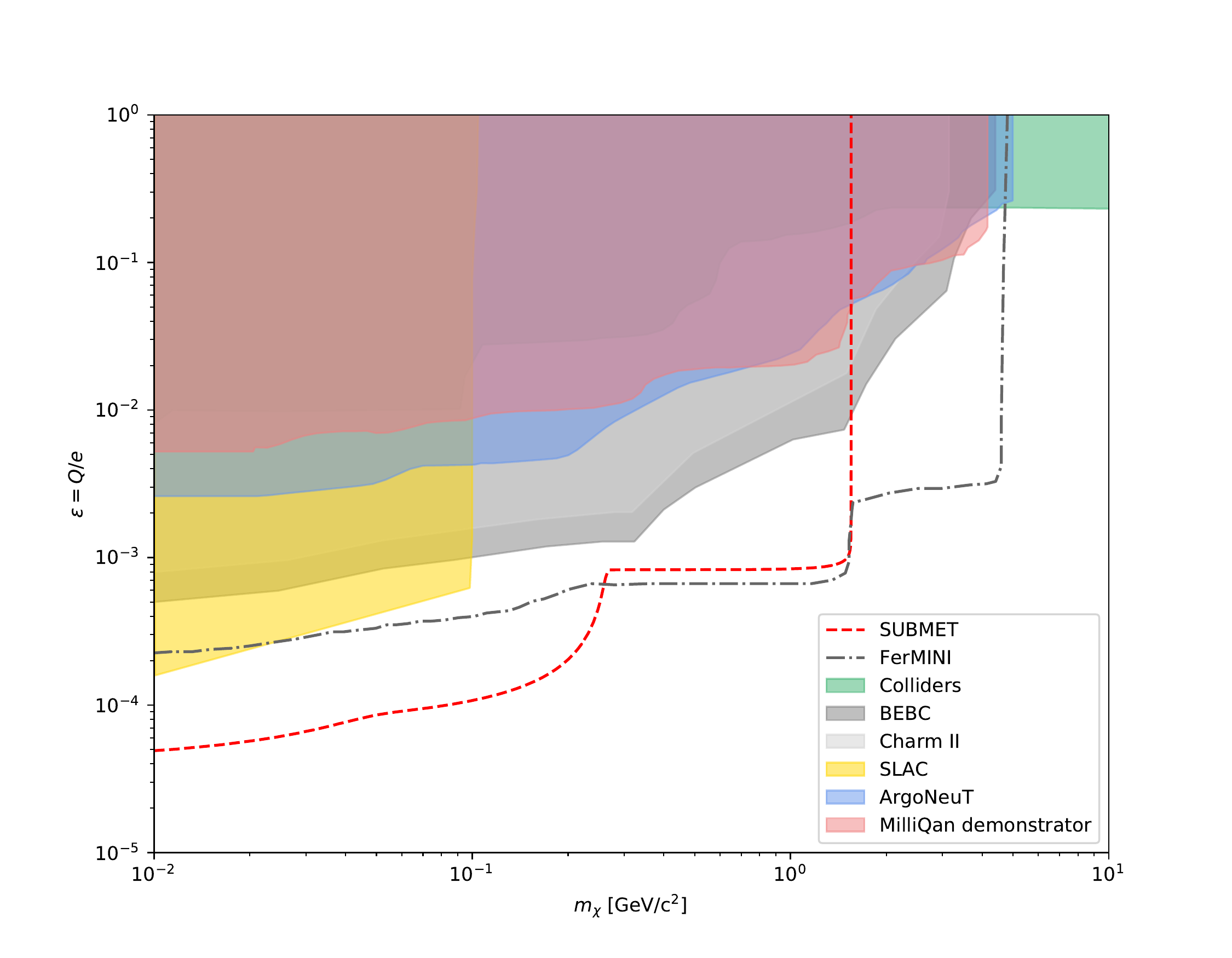}
\end{center}
\caption{Exclusion at $95$\% CL for $N_\textrm{POT}=10^{22}$. The constraints from previous experiments are shown as shaded areas. The expected sensitivity of FerMINI~\cite{fermini} is drawn in the gray dotted line. There are other proposed experiments~\cite{Ball:2016zrp,Harnik:2019zee}, but only FerMINI with the NuMI beam is included because it is in a similar time scale of SUBMET (within next 5 years).}
\label{fig:limit}
\end{figure}

Figure~\ref{fig:limit} shows the $95$\% CL exclusion curve for $N_\textrm{POT}=10^{22}$. SUBMET provides the exclusion down to $\epsilon=5\times10^{-5}$ in $m_\chi<0.2$ $\textrm{GeV}/\textrm{c}^2$ and $\epsilon=8\times10^{-4}$ in $m_\chi<1.6$ $\textrm{GeV}/\textrm{c}^2$. Systematic uncertainty on $b$ is not considered because it does not have a significant impact on the exclusion limit (Table~\ref{tab:det_config}). The sudden degradation of sensitivity at $m_\chi=0.2$$\textrm{GeV}/\textrm{c}^2$ is because of the small production rate of $J/\psi$ with the $30$ GeV proton beam.

The number of signal events recorded by the detector drops rapidly in $\epsilon<10^{-3}$ due to small $N_\textrm{PE}$. Therefore, increasing $N_\textrm{PE}$ or $N_\chi$ does not have a large impact on the sensitivity in this phase space. This will be discussed in a quantitative way in the next section.

\section{Alternative detector design}
\label{sec:alterdetector}

The sensitivity of the experiment depends on the configuration of the detector. This section describes the impact of a few key parameters for the detector design, focusing on the sub-millicharge regime. If further optimization of the detector is needed, this quantitative study can serve as a guide.


Searches in the sub-millicharge regime rely on the Poissonian fluctuation of small $N_\textrm{PE}$. Approximating $P \simeq N_\textrm{PE}$ to the first order for small $\epsilon$, we arrive at the following relation
\begin{eqnarray}
\label{eq:s95}
s = N_{\epsilon=1,\chi} \epsilon^2 (N_{\epsilon=1,\textrm{PE}} \epsilon^2)^n \geq s^{\textrm{95\%}}
\end{eqnarray}
where $s$ is the number of signal events, the subscript $\epsilon = 1$ refers to the values at $\epsilon = 1$ and $s^{\textrm{95\%}}$ is the number of signal events that provides 95\% exclusion limit. Reordering in terms of $\epsilon$, the exclusion limit at 95\% CL is
\begin{eqnarray}
\label{eq:eps95}
\epsilon = \left( \frac{N_{\epsilon=1, \textrm{PE}}^n N_{\epsilon=1,\chi}}{s^{\textrm{95\%}}}\right)^{-\frac{1}{2n+2}}.
\end{eqnarray}
Due to the $\epsilon^{2(n+1)}$ term in~\ref{eq:s95}, there is a sharp cutoff in $s$ around $\epsilon \sim O(10^{-6})$. This limits the sensitivity to that regime regardless of the detector configuration.

\begin{table}
  \begin{center}
    \begin{tabular}{c c c | c}
      \hline
      \multirow{2}{*}{$N_\chi$ (relative)}  &  \multirow{2}{*}{$N_\textrm{PE}$ (relative)} & \multirow{2}{*}{$b$ (relative)} &  Exclusion limit on $\epsilon$ \\
       &  &  &  for $m_{\chi} = 10$ $\textrm{MeV}/\textrm{c}^2$\\
      \hline\hline
      $1$ & $1$ & $1$ & $4.9\times10^{-5}$\\
      \hline
      $2$ & $1$ & $1$ & $4.6\times10^{-5}$\\
      \hline
      $1/50$ & $1$ & $1$ & $8.4\times10^{-5}$\\
      \hline
      $1$ & $1$ & $25$ & $6.0\times10^{-5}$\\
      \hline
      $1$ & $2$ & $1$ & $3.9\times10^{-5}$\\
      \hline
      $1$ & $0.8$ & $1$ & $5.3\times10^{-5}$\\
      \hline
    \end{tabular}
  \end{center}
  \caption{Various detector configurations and their sensitivity. $N_\chi$ (relative) is the yields of signal events within the acceptance relative to the baseline, $N_\textrm{PE}$ (relative) is the number of photon electrons relative to $N_\textrm{PE}=2.5\times10^5$, and $b$ (relative) is the number of background events relative $N_\textrm{bkg}=5$.}
 \label{tab:det_config}
\end{table}

Table~\ref{tab:det_config} shows different detector configurations and the corresponding exclusion limits for $m_{\chi} = 10$ $\textrm{MeV}/\textrm{c}^2$. The default configuration is in the first row; $N_\chi$(relative)=1, $N_\textrm{PE}$(relative)=1, and $b$(relative) $=1$ where ``relative'' means relative to the values of the baseline configuration discussed in Section~\ref{sec:site_detector}.

The improvements achieved by extending the duration of data collection or making the detector area larger, \textit{e.g.}, adding more bars to each layer, are modest. Extending the duration or the detector area by a factor of 2 increases $N_\chi$ by the same amount. Sensitivity is improved by less than $10$\% ($2^\textrm{nd}$ row in Table~\ref{tab:det_config}). If the area of the detector is reduced to $10\times10$ cm (a factor of $50$) or the duration of the data-taking period is shortened by $1/50$ (roughly 3 weeks), the exclusion limit moves to $8.4\times10^{-5}$ ($3^\textrm{rd}$ row in Table~\ref{tab:det_config}). The impact of $b$ is limited as well. If $b$ is increased by a factor of $25$ which corresponds to the DCR of $500$ Hz, the limit is degraded by only $20$\% ($4^\textrm{th}$ row in Table~\ref{tab:det_config}).

The $5^\textrm{th}$ row in Table~\ref{tab:det_config} shows that the most effective component to enhance sensitivity is $N_\textrm{PE}$. Increasing $N_\textrm{PE}$ by a factor of $2$ improves the sensitivity by $20$\%. This can be achieved by using a scintillator material with a higher light yield or extending the length of the scintillator. However, improvement using longer scintillators is limited in length $>150$ cm due to the effect of light attenuation. The $6^\textrm{th}$ row corresponds to the case of scintillator length of $75$ cm. Degradation of sensitivity is only 10 \%, which allows for a smaller-scale detector without a significant loss of performance. 

Though $N_\textrm{PE}$ plays the main role in enhancing sensitivity to sub-millicharged particles, the exclusion limit with the $0.5\times0.5$ $\textrm{ m}^2$ detector is still in the range of $\epsilon=(3.9-6.0)\times10^{-5}$ with the variations considered. This indicates that the sensitivity does not depend on the configuration of the detector strongly.

In case of an unexpectedly large number of background events, installing additional layers can be considered to control them. Using 3 layers and assuming 0 background events, the exclusion limit reaches $\epsilon=1.1\times10^{-4}$. The experiment still outperforms previous searches in this configuration. 


\section{Discussion and Conclusion}

We propose a new experiment, SUBMET, sensitive to millicharged particles produced at the $30$ GeV proton fixed-target collisions at J-PARC. The detector, inspired by the milliQan experiment, is based on long scintillators and is located in the Neutrino Monitor building 280 m from the target. With the number of protons on target of $10^{22}$, the experiment is sensitive to particles with electric charge $5\times10^{-5}\,e$ for mass less than $0.2\textrm{ GeV}/\textrm{c}^2$ and $8\times10^{-4}\,e$ for mass less than $1.6$ $\textrm{GeV}/\textrm{c}^2$. 

SUMBET places the best limit in low mass region $m_\chi<0.2\textrm{ GeV}/\textrm{c}^2$ among the existing and the proposed experiments. In this regime, the $N_{PE}$ is very small so the probability to observe a photon produced by millicharged particles per layer ($P_\textrm{layer}=(1-e^{-N_\textrm{PE}}$)) is extremely small. Since the total probability is a power of $P_\textrm{layer}$ by the number of layers, using two layers significantly enhances the probability compared to the detector designs with 3 or 4 layers. 

Note that this experiment is complementary to the existing proposals~\cite{Ball:2016zrp,fermini,formosa} since the main interest is in the low mass region. The center of mass energy of the proton-target collisions is $7.5$ GeV and this limits the mass reach of the experiment to below $m_{J/\psi}/2$ while other proposals can cover higher mass regions. Compared to the FerMINI experiment, the production rate of $J/\psi$ is much smaller due to lower beam energy. So, the sensitivity to the $\chi$s from $J/\psi$ decay is slightly worse though it is still competitive.   


A few detector designs to achieve an optimal sensitivity were considered in Section~\ref{sec:alterdetector} and we found that the configuration of the detector generally does not affect the sensitivity. In addition, the operation of the upgraded proton beam at J-PARC will start in early 2022~\cite{Friend_2017}. These indicate that it is very important to install the detector as early as possible to fully exploit the upgraded power of the beam.        


\acknowledgments

Authors thank Tsutomu Mibe, Yoshiaki Fujii, Takeshi Nakadaira, and Toshifumi Tsukamoto for the useful discussions on the detector site. In particular, we thank Toshifumi Tsukamoto for taking photographs of the Neutrino Monitor building so that we understand the spatial constraints inside the building. We thank the members of the milliQan collaboration, particularly, Andy Hass, Christopher S. Hill, and David Stuart for the discussions at various stages of this study. We thank Matthew Citron, Albert De Roeck, Seung Joon Lee, and Eunil Won for providing comments on the draft. We thank Masashi Yokoyama for discussions regarding the schedule of the neutrino beamline. We also thank Hong Joo Kim for the information on the property of various scintillation materials. This work has been supported by a Korea University Grant.



\end{document}